\documentclass[12pt,a4paper]{article}

\usepackage[utf8]{inputenc}
\usepackage[T1]{fontenc}
\usepackage[brazil]{babel}
\usepackage{amsmath, amssymb}
\usepackage{graphicx}
\usepackage{booktabs}
\usepackage{geometry}
\usepackage{natbib}
\usepackage{subcaption}
\usepackage{csquotes}
\usepackage{longtable}
\usepackage{xurl}
\usepackage{hyperref}
\usepackage{float}
\usepackage{setspace}
\usepackage{titling}
\usepackage{microtype}

\hypersetup{
    colorlinks=true,
    linkcolor=blue,
    urlcolor=blue,
    citecolor=blue,
    breaklinks=true
}

\title{Como medir o invisível? \\ Guerras, pizzarias do Pentágono e o uso de variáveis \emph{proxy} em econometria}

\author{
    {\large Guilherme Vianna}\thanks{\href{mailto:guilherme.dias.vianna@usp.br}{guilherme.dias.vianna@usp.br}} \\
    {\normalsize USP}
    \and
    {\large Victor Rangel}\thanks{\href{mailto:victorrsr@al.insper.edu.br}{victorrsr@al.insper.edu.br}} \\
    {\normalsize Insper}
}

\date{}

\begin{document}
\maketitle

\section*{Introdução}

Muitas variáveis importantes para entender o mundo econômico não vêm em séries oficiais: qualidade institucional, confiança do consumidor, custo de fiscalização, até o humor dos investidores. Não existe uma planilhona do FMI \texttt{RiscoFiscal.csv} com todos os países e um número definitivo dado pelos deuses. Essas quantidades, porém, existem, afetam decisões e preços, mas são latentes: não as observamos diretamente. Como conciliá-las em econometria? A saída clássica é usar variáveis \emph{proxy}: medidas observáveis que reagem ao mesmo ``impulso'' do que queremos medir, permitindo inferir o invisível.

A ``teoria da pizza'' também ganhou uma camada contemporânea: passou a ser monitorada em tempo real por contas e painéis de internet, e voltou a circular com força em eventos específicos. Em junho de 2025, reportagens e postagens destacaram um pico de atividade em pizzarias próximas ao Pentágono pouco antes de um episódio de escalada entre Israel e Irã, tratado por muitos como um sinal informal de ``noite longa'' no aparato de defesa.\footnote{Uma reportagem popularizando essa leitura: \url{https://veja.abril.com.br/mundo/como-delivery-de-pizzas-para-o-pentagono-previu-ataque-de-israel-ao-ira/}. Para uma visão mais cética e contextualizada sobre a dinâmica do monitoramento, ver também: \url{https://www.washingtonpost.com/food/2025/07/01/pentagon-pizza-tracker-orders-military/}.} Já no início de 2026, o mesmo tipo de monitoramento voltou a viralizar em torno da operação anunciada como culminando na captura de Nicolás Maduro.\footnote{Coberturas do episódio e da associação com o ``Índice da Pizza'': \url{https://exame.com/mundo/indice-da-pizza-mostra-pico-na-madrugada-em-que-maduro-foi-capturado/} e \url{https://www.cnnbrasil.com.br/internacional/entenda-o-que-e-o-indice-da-pizza-e-a-relacao-com-o-ataque-a-venezuela/}.} 

Então, para fixar as ideias, considere o caso das pizzarias ao redor do Pentágono, como indicado na Figura \ref{fig:pizza}:

\begin{figure}[H]
  \centering
  \includegraphics[width=.5\linewidth]{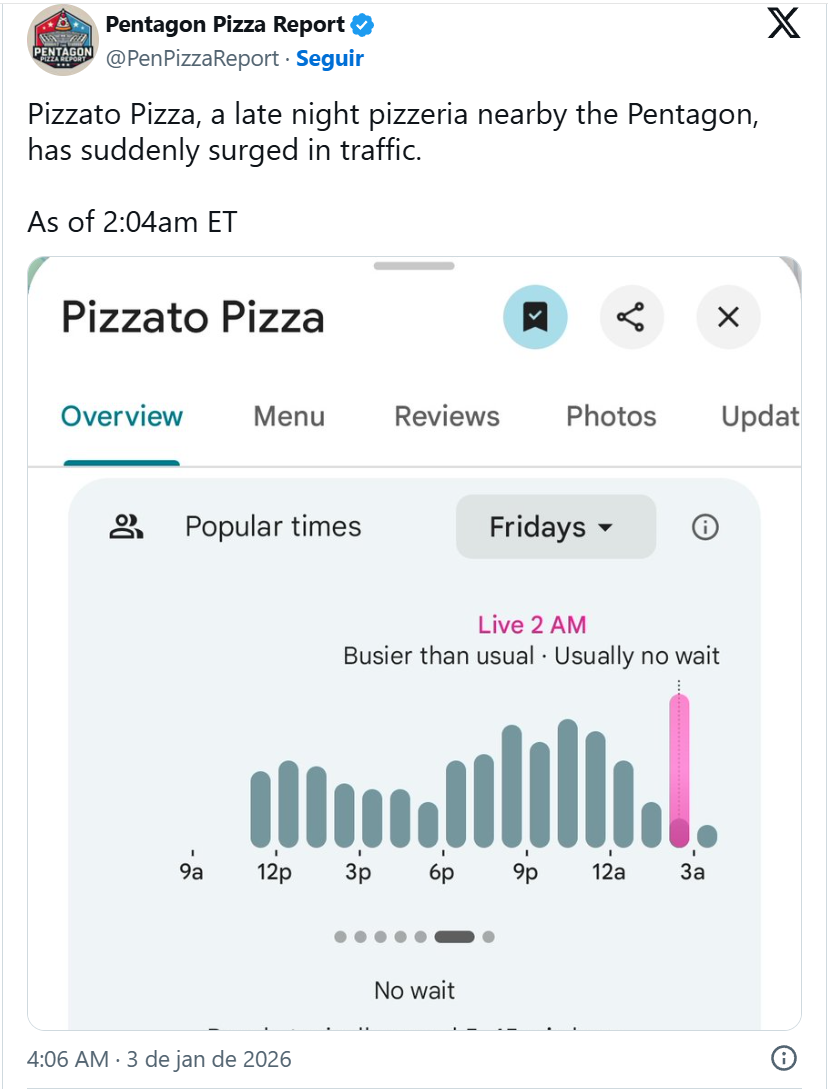}
  \caption{Pizza Index: Excesso de demanda em pizzaria próxima ao Pentágono}\label{fig:pizza}
\end{figure}

Em momentos de tensão geopolítica, equipes estendem jornada; os pedidos de pizza na vizinhança disparam. A aposta é que o volume de pedidos seja um termômetro do nível de crise. Se isso for verdade, essa \emph{proxy} poderia ajudar a entender como a crise mexe com ações da indústria bélica, petróleo, volatilidade dos mercados ou o risco geopolítico do país. \footnote{Tanto os exemplos quanto a aplicação empírica propostas são usados exclusivamente como recursos didáticos para ilustrar a lógica de variáveis \emph{proxy}, e a passagem da teoria para a prática. Os resultados não são apresentados como descobertas empíricas definitivas nem como evidência causal no sentido estrito. Nosso objetivo é pedagógico: oferecer uma ponte clara entre intuição, formalização e um \emph{toy example} replicável em sala de aula.}

\section*{O modelo verdadeiro}

Imagine que, por simplicidade, temos o seguinte modelo:
\begin{equation}
Y = \alpha_0 + \alpha_1 X + \alpha_2 C + U,  \label{eq:modelo_verdadeiro}
\end{equation}
onde $Y$ é o preço de uma ação de alguma empresa da indústria bélica (por exemplo, Lockheed Martin); $X$ indica um índice amplo de mercado (observável), como o S\&P500; $C$ é o nível de crise ou de incerteza geopolítica (latente, não observável); e $U$ agrega choques externos idiossincráticos e imprevisíveis, com $\mathbb{E}[U\mid X, C] = 0$. Sob as hipóteses usuais (linearidade, exogeneidade, homoscedasticidade, ausência de autocorrelação e sem colinearidade perfeita), o MQO identificaria bem \(\alpha_1\) e \(\alpha_2\).

\section*{Se omitimos $C$: o viés de variável omitida aparece}

Na prática, como $C$ não é observável, alguém poderia estimar a regressão reduzida
\[
Y = \beta_0 + \beta_1 X + \varepsilon.
\]

O que ela obteria?

Dada uma amostra observada, a estimativa de MQO com um regressor é
\[
\hat\beta_1 = \frac{\widehat{\mathbb{C}\mathrm{ov}}(X, Y)}{\widehat{\mathbb{V}}(X)} = \frac{\widehat{\mathbb{C}\mathrm{ov}}(X, \alpha_0 + \alpha_1 X + \alpha_2 C + U)}{\widehat{\mathbb{V}}(X)}  = \alpha_1 + \alpha_2 \frac{\widehat{\mathbb{C}\mathrm{ov}}(X, C)}{\widehat{\mathbb{V}}(X)} + \frac{\widehat{\mathbb{C}\mathrm{ov}}(X, U)}{\widehat{\mathbb{V}}(X)}.
\]

Veja que:
\[
\mathbb{E}[\hat\beta_1 \mid X, C] = \mathbb{E}\left[ \alpha_1 + \alpha_2 \frac{\widehat{\mathbb{C}\mathrm{ov}}(X, C)}{\widehat{\mathbb{V}}(X)} + \frac{\widehat{\mathbb{C}\mathrm{ov}}(X, U)}{\widehat{\mathbb{V}}(X)} \Bigg| X, C \right]
\]
\[
= \alpha_1 + \mathbb{E}\left[\alpha_2 \frac{\widehat{\mathbb{C}\mathrm{ov}}(X, C)}{\widehat{\mathbb{V}}(X)} \Bigg| X,C \right] + \mathbb{E}\left[\frac{\widehat{\mathbb{C}\mathrm{ov}}(X, U)}{\widehat{\mathbb{V}}(X)} \Bigg| X,C \right].
\]
Como os dois primeiros termos são funções de \(X\) e \(C\), eles são tratados como constantes pela esperança condicional. O último termo é zero pela hipótese de $\mathbb{E}[U\mid X,C]=0$. Assim,
\[
\mathbb{E}[\hat\beta_1 \mid X, C] = \alpha_1 + \alpha_2 {\frac{\widehat{\mathbb{C}\mathrm{ov}}(X, C)}{\widehat{\mathbb{V}}(X)}}.
\]
Logo,
\[
\text{Viés} = \mathbb{E}[\hat\beta_1 \mid X, C] - \alpha_1 = \alpha_2 {\frac{\widehat{\mathbb{C}\mathrm{ov}}(X, C)}{\widehat{\mathbb{V}}(X)}}.
\]

Veja que temos um termo além do que gostaríamos: o segundo termo é o viés de variável omitida. Se $X$ costuma subir quando $C$ aumenta e se $C$ puxa $Y$ para cima, então o efeito de $X$ ficará inflado. Se os sinais forem opostos, ficará subestimado e até mesmo com o sinal contrário. Ou seja, a presença de uma variável não observável influenciando \(Y\) não é só mera inconveniência: omiti-la da regressão faz com que você estime de forma errada a influência das outras variáveis.

\section*{Incluindo a \emph{proxy}: pizzas como $P$}

Suponha agora que o número de pedidos de pizza numa determinada semana (ou algo do gênero) é diretamente proporcional ao nível da crise (uma \emph{proxy perfeita}):
\[
P = \lambda C,\quad \lambda > 0.
\]
Considere então que eu queira ajustar o seguinte modelo com dois regressores observáveis:
\[
Y = \gamma_0 + \gamma_1 X + \gamma_2 P + E.
\]

Será que isso resolve o problema do viés?

A fórmula do estimador de MQO para \(\gamma_1\) na presença de um segundo regressor e um intercepto é:
\[
\hat\gamma_1 = \frac{\widehat{\mathbb{C}\mathrm{ov}}(X, Y) \widehat{\mathbb{V}}(P) - \widehat{\mathbb{C}\mathrm{ov}}(P, Y) \widehat{\mathbb{C}\mathrm{ov}}(X, P)}{\widehat{\mathbb{V}}(X) \widehat{\mathbb{V}}(P) - [\widehat{\mathbb{C}\mathrm{ov}}(X, P)]^2}.
\]

Usando o modelo completo \(Y = \alpha_0 + \alpha_1 X + (\alpha_2/\lambda)P + U\), as covariâncias são:
\[
\widehat{\mathbb{C}\mathrm{ov}}(X, Y)   = \alpha_1 \widehat{\mathbb{V}}(X) + \frac{\alpha_2}{\lambda} \widehat{\mathbb{C}\mathrm{ov}}(X, P) + \widehat{\mathbb{C}\mathrm{ov}}(X, U),
\]
\[
\widehat{\mathbb{C}\mathrm{ov}}(P, Y) = \alpha_1 \widehat{\mathbb{C}\mathrm{ov}}(P, X) + \frac{\alpha_2}{\lambda} \widehat{\mathbb{V}}(P) + \widehat{\mathbb{C}\mathrm{ov}}(P, U).
\]
Substituindo estas expressões no numerador de \(\hat\gamma_1\), os termos com \(\alpha_2/\lambda\) se cancelam, e os termos com \(\alpha_1\) se agrupam:
\[
\text{Numerador} = \alpha_1 \left( \widehat{\mathbb{V}}(X)\widehat{\mathbb{V}}(P) - [\widehat{\mathbb{C}\mathrm{ov}}(X, P)]^2 \right) + \left( \widehat{\mathbb{C}\mathrm{ov}}(X, U)\widehat{\mathbb{V}}(P) - \widehat{\mathbb{C}\mathrm{ov}}(P, U)\widehat{\mathbb{C}\mathrm{ov}}(X, P) \right).
\]
Dividindo pelo denominador, a expressão para o estimador se torna:
\[
\hat\gamma_1 = \alpha_1 + \frac{\widehat{\mathbb{C}\mathrm{ov}}(X, U)\widehat{\mathbb{V}}(P) - \widehat{\mathbb{C}\mathrm{ov}}(P, U)\widehat{\mathbb{C}\mathrm{ov}}(X, P)}{\widehat{\mathbb{V}}(X)\widehat{\mathbb{V}}(P) - [\widehat{\mathbb{C}\mathrm{ov}}(X, P)]^2}.
\]
O valor esperado do segundo termo condicional em $X$ e $P$ é zero, sob a hipótese de que \(\mathbb{E}[U\mid X,P]=0\). Portanto:
\[
\mathbb{E}[\hat\gamma_1 \mid X, P] = \alpha_1.
\]

Ou seja, no caso de uma \emph{proxy} perfeita, é possível limpar o viés de variável omitida perfeitamente.\footnote{O nome \emph{proxy perfeita} não é à toa.} E como será que a nossa estimativa de \(\gamma_2\) se relacionará com \(\alpha_2\)?

\[
\hat\gamma_2 = \frac{\widehat{\mathbb{C}\mathrm{ov}}(P, Y) \widehat{\mathbb{V}}(X) - \widehat{\mathbb{C}\mathrm{ov}}(X, Y) \widehat{\mathbb{C}\mathrm{ov}}(X, P)}{\widehat{\mathbb{V}}(X) \widehat{\mathbb{V}}(P) - [\widehat{\mathbb{C}\mathrm{ov}}(X, P)]^2}.
\]
Usando as mesmas expressões para as covariâncias com \(Y\) derivadas anteriormente, agora os termos com \(\alpha_1\) se cancelam, e os termos com \(\alpha_2/\lambda\) se agrupam:
\[
\text{Numerador} = \frac{\alpha_2}{\lambda} \left( \widehat{\mathbb{V}}(X)\widehat{\mathbb{V}}(P) - [\widehat{\mathbb{C}\mathrm{ov}}(X, P)]^2 \right) + \left( \widehat{\mathbb{C}\mathrm{ov}}(P, U)\widehat{\mathbb{V}}(X) - \widehat{\mathbb{C}\mathrm{ov}}(X, U)\widehat{\mathbb{C}\mathrm{ov}}(X, P) \right).
\]
Dividindo pelo denominador, a expressão para o estimador se torna:
\[
\hat\gamma_2 = \frac{\alpha_2}{\lambda} + \frac{\widehat{\mathbb{C}\mathrm{ov}}(P, U)\widehat{\mathbb{V}}(X) - \widehat{\mathbb{C}\mathrm{ov}}(X, U)\widehat{\mathbb{C}\mathrm{ov}}(X, P)}{\widehat{\mathbb{V}}(X)\widehat{\mathbb{V}}(P) - [\widehat{\mathbb{C}\mathrm{ov}}(X, P)]^2}.
\]
O valor esperado do segundo termo condicional em \(X\) e \(P\) é zero, sob a hipótese de que \(\mathbb{E}[U\mid X,P]=0\). Portanto:
\[
\mathbb{E}[\hat\gamma_2 \mid X, P] = \frac{\alpha_2}{\lambda}.
\]

Portanto, a inclusão da \emph{proxy} perfeita resolve o problema de viés e nos leva a duas conclusões práticas, a depender do nosso conhecimento sobre a escala \(\lambda\). Se \(\lambda\) for conhecido (por teoria, estudos anteriores ou pela própria construção da \emph{proxy}), podemos criar um estimador \(\hat\alpha_{2}^{\text{proxy}} = \lambda \cdot \hat\gamma_2\), não enviesado para \(\alpha_2\). Contudo, na situação mais comum em que \(\lambda\) é desconhecido, o resultado ainda é valioso: como \(\mathbb{E}[\hat\gamma_2] = \alpha_2/\lambda\) e \(\lambda > 0\), ainda conseguimos determinar corretamente o sinal de \(\alpha_2\) e realizar testes de hipótese válidos para avaliar sua significância estatística.

\subsection*{Quando a \emph{proxy} é boa, mas não perfeita}

Mais realista é admitir que $P$ não seja exatamente proporcional a $C$ e que haja algum ruído na relação entre elas, algo da forma:
\[
C = \delta_0 + \delta_X X + \delta_P P + V,
\]
com $V$ sendo o ``resto'' não previsível por $X$ e $P$ (isto é, $\mathbb{E}[V\mid X,P]=0$). Substituindo isso no modelo original (Eq.~\ref{eq:modelo_verdadeiro}), obtemos:
\[
Y \;=\; (\alpha_0 + \alpha_2 \delta_0) \;+\; (\alpha_1 + \alpha_2 \delta_X)\,X \;+\; (\alpha_2 \delta_P)\,P \;+\; (\alpha_2 V + U).
\]
Ou, reescrevendo:
\[
Y \;=\; \mu_0 \;+\; \mu_X X \;+\; \mu_P P \;+\; E,\footnote{Equivalências:
$\mu_0 \equiv \alpha_0 + \alpha_2 \delta_0$,
$\mu_X \equiv \alpha_1 + \alpha_2 \delta_X$,
$\mu_P \equiv \alpha_2 \delta_P$,
$E \equiv \alpha_2 V + U$ com $\mathbb{E}[E\mid X,P]=0$.}
\]

Como ficam os estimadores desses novos parâmetros?

As fórmulas são as mesmas, mas agora a expressão de \(C\) muda. O resultado final é:
\[
\hat\mu_X = \mu_X + \frac{\widehat{\mathbb{C}\mathrm{ov}}(X, E)\widehat{\mathbb{V}}(P) - \widehat{\mathbb{C}\mathrm{ov}}(P, E)\widehat{\mathbb{C}\mathrm{ov}}(X, P)}{\widehat{\mathbb{V}}(X)\widehat{\mathbb{V}}(P) - [\widehat{\mathbb{C}\mathrm{ov}}(X, P)]^2}.
\]
Tomando a esperança condicional:
\[
\mathbb{E}[\hat\mu_X \mid X, P] = \mu_X = \alpha_1 + \alpha_2 \delta_X.
\]
A análise revela que nosso estimador para o efeito do mercado, \(\hat\mu_X\), carrega um viés residual de \(\alpha_2 \delta_X\). A questão central, portanto, se desloca para a plausibilidade da hipótese de que \(\delta_X = 0\). Essa é uma suposição forte: ela afirma que o S\&P500 (\(X\)) não contém nenhuma informação relevante para explicar a incerteza geopolítica (\(C\)) que já não esteja sendo capturada pela \emph{proxy}, os pedidos de pizza (\(P\)).

No contexto do nosso exemplo, essa hipótese é defensável? É aqui que a teoria e o conhecimento institucional se tornam indispensáveis. Pode-se argumentar que, em tempos normais, a causalidade principal flui da incerteza geopolítica \emph{para} o mercado, e não o contrário. Flutuações cotidianas do S\&P500 provavelmente não geram crises internacionais. Sob essa visão, seria razoável supor \(\delta_X \approx 0\) no modelo populacional.

Contudo, a relação pode não ser estável. Em cenários de pânico financeiro ou de uma recessão global, uma queda drástica do mercado poderia agravar tensões e instabilidade, tornando a fragilidade econômica um estopim para conflitos. Nesse caso, \(\delta_X\) seria diferente de zero e o viés ressurgiria.

O ponto crucial é que a decisão de assumir \(\delta_X = 0\) não pode ser validada empiricamente na amostra: tentar ``testar'' se \(X\) se correlaciona com \(C\) é impossível nesta configuração, pois \(C\) é inobservável. Trata-se de uma hipótese de identificação sobre o funcionamento do mundo, que o econometrista deve justificar com base em argumentação teórica e conhecimento institucional.

Voltando, o resultado para o estimador \(\hat{\mu}_P\) é:
\[
\hat\mu_P = \mu_P + \frac{\widehat{\mathbb{C}\mathrm{ov}}(P, E)\widehat{\mathbb{V}}(X) - \widehat{\mathbb{C}\mathrm{ov}}(X, E)\widehat{\mathbb{C}\mathrm{ov}}(X, P)}{\widehat{\mathbb{V}}(X)\widehat{\mathbb{V}}(P) - [\widehat{\mathbb{C}\mathrm{ov}}(X, P)]^2}.
\]
Tomando a esperança condicional:
\[
\mathbb{E}[\hat\mu_P \mid X, P] = \mu_P = \alpha_2 \delta_P.
\]

O coeficiente da \emph{proxy} estima o efeito verdadeiro \(\alpha_2\) atenuado pela qualidade da \emph{proxy}, \(\delta_P\). Esse resultado amarra dois pontos da identificação: o ``quanto'' a \emph{proxy} informa sobre o fator latente depende de \(\delta_P\), e o ``quão limpo'' fica o coeficiente da variável de interesse depende de assumir \(\delta_X=0\).

\section*{O que esperamos de uma boa \emph{proxy}, mesmo que imperfeita?}

A partir daqui, adotaremos um protocolo prático para avaliar \emph{proxies} no mundo real. A ideia é separar claramente quatro propriedades que, na prática, precisam ser defendidas com argumento econômico e conhecimento institucional.

\paragraph{Relevância.}
A \emph{proxy} \(P\) precisa ter correlação com a variável latente \(C\), mesmo após controlar pelos outros regressores. Na nossa projeção, essa condição é traduzida por \(\delta_P\). Queremos \(\delta_P\) diferente de zero e, idealmente, grande em magnitude. Como \(\mathbb{E}[\hat\mu_P] = \alpha_2 \delta_P\), se \(\delta_P\) for pequeno o efeito estimado da \emph{proxy} será muito atenuado (próximo de zero), mesmo que o efeito verdadeiro \(\alpha_2\) seja grande.

\paragraph{Suficiência condicional.}
Esta é a hipótese crucial para eliminar o viés no coeficiente da variável de interesse, \(\alpha_1\). A condição diz que, uma vez que conhecemos o valor da \emph{proxy} \(P\), a outra variável \(X\) não deve conter nenhuma informação adicional sobre a variável omitida \(C\). Matematicamente, na projeção
\[
\mathbb{E}[C \mid X, P] = \delta_0 + \delta_X X + \delta_P P,
\]
a suficiência condicional requer \(\delta_X = 0\). Quando isso ocorre, o viés em \(\hat\mu_X\), que era \(\alpha_2 \delta_X\), desaparece.

\paragraph{Exogeneidade.}
Esta condição se refere à relação entre a \emph{proxy} e o erro estrutural \(U\) do modelo original. A hipótese \(\mathbb{E}[U \mid X,P]=0\) afirma que a \emph{proxy} \(P\) só afeta \(Y\) através de sua correlação com \(C\). No nosso exemplo, isso significa que um aumento nos pedidos de pizza (\(P\)) não pode ter um efeito direto no preço das ações (\(Y\)) por canais próprios que não passem pela incerteza geopolítica (\(C\)).

\paragraph{Estabilidade.}
Esta propriedade requer que a relação entre a \emph{proxy}, a variável latente e os outros regressores seja aproximadamente constante ao longo do tempo. Em termos dos parâmetros, isso significa que \(\delta_0\), \(\delta_X\) e \(\delta_P\) não sofrem quebras estruturais relevantes. Se a maneira como \(C\) se manifesta em \(P\) mudar drasticamente no meio da análise, a interpretação dos coeficientes se torna frágil.

A econometria aplicada frequentemente se depara com um dilema: muitas das forças mais interessantes que moldam o comportamento econômico --- como risco, confiança ou incerteza --- são, por natureza, invisíveis aos dados. Como vimos, a simples omissão dessas variáveis latentes não é uma opção neutra; ela distorce nossa compreensão das relações observáveis. As variáveis \emph{proxy} surgem, então, como uma ferramenta para dar contorno a essas forças, permitindo que o invisível deixe sua marca em nossos modelos.

\section*{Um exemplo aplicado: micromobilidade em Arlington como \emph{proxy} de risco geopolítico nos EUA}

A jornada de uma \emph{proxy} perfeita para uma imperfeita é a transição da teoria para a prática. Na ausência de um termômetro perfeito para a ``crise geopolítica'', medidas alternativas oferecem retratos ruidosos, mas úteis. A análise formal mostra que, ao usar uma \emph{proxy} imperfeita, trocamos um viés de magnitude desconhecida por um viés residual, \(\alpha_2\delta_X\), agora explicitado e discutível; em paralelo, o coeficiente associado à \emph{proxy}, \(\alpha_2\delta_P\), embora atenuado, informa direção e relevância do fator omitido. O progresso não está em obter a resposta ``certa'' de imediato, mas em caracterizar com precisão a incerteza remanescente. Para tornar essa intuição observável, levamos a analogia das ``pizzarias do Pentágono'' ao mundo real usando dados de micromobilidade no entorno do Pentágono.

\begin{figure}[H]
  \centering
  \includegraphics[width=.8\linewidth]{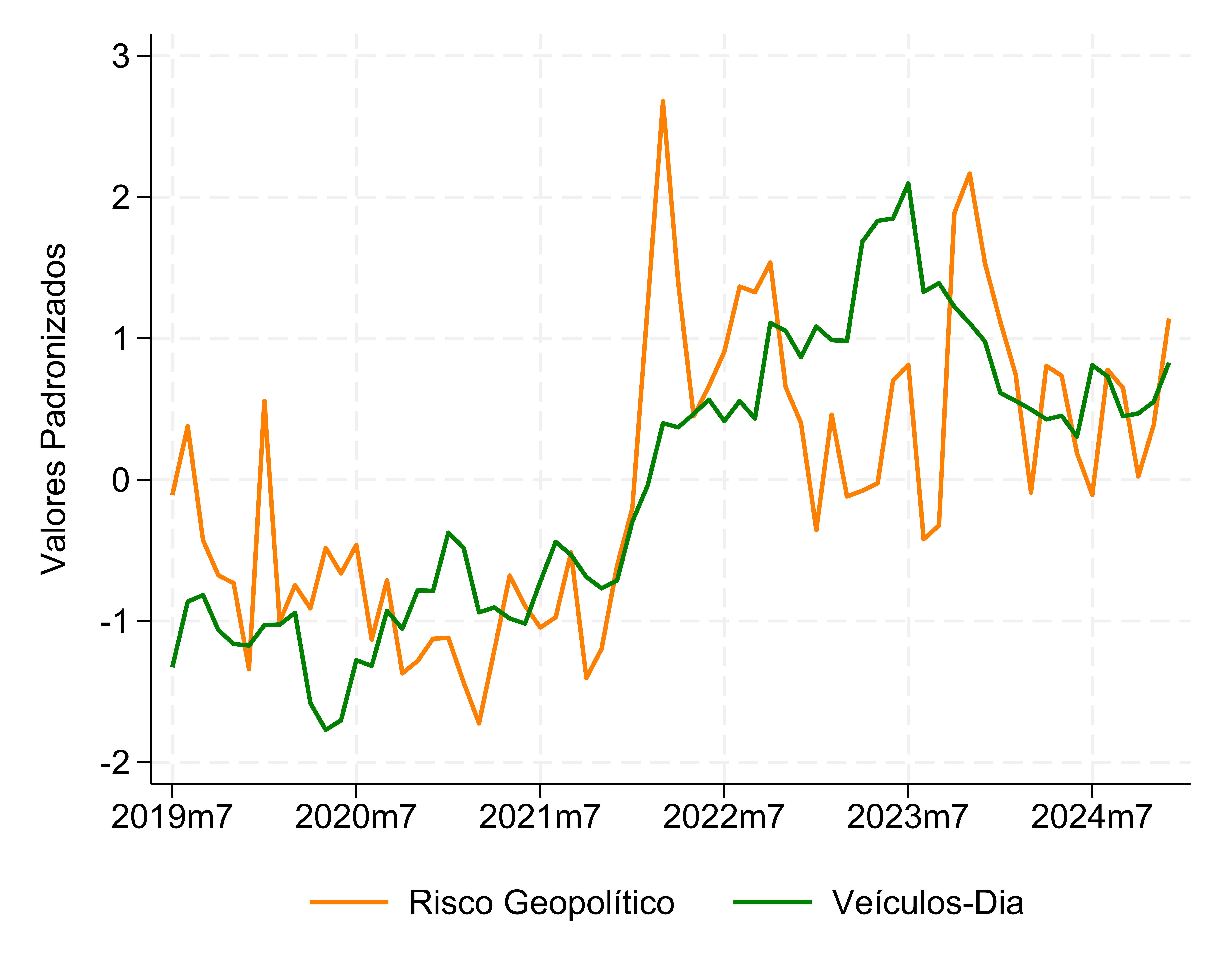}
  \caption{Relação entre Risco Geopolítico e Micromobilidade}\label{fig:comparacao}
\end{figure}

A analogia sugere que rotinas locais ao redor dos centros de decisão revelam, indiretamente, o ritmo das tensões geopolíticas. Implementamos essa ideia com a série de micromobilidade de Arlington (\emph{All Vehicle Trips per Day}\footnote{Fonte: \url{https://public.ridereport.com/}.}) e o índice nacional de risco geopolítico dos EUA (GPR) \citep{caldara2022measuring}. A Figura~\ref{fig:comparacao} ilustra a coevolução dessas séries.

A motivação econômica é direta: quando a agenda de defesa ``esquenta'', aumentam jornadas, pedidos de \emph{delivery}, reuniões e deslocamentos de curta distância no entorno institucional; tarefas urgentes de última milha e \emph{micro-commutes} migram para meios rápidos e flexíveis (scooters e bikes compartilhadas), gerando um sinal denso, de alta frequência e geograficamente preciso no ponto de origem do impulso. Assim, a micromobilidade registra variações do mesmo fator subjacente que move o GPR, sem pretender causá-lo: em tempos mais tensos, espera-se mais atividade local e, portanto, co-movimento entre viagens de micromobilidade e o risco geopolítico agregado, que é um racional claro para tratá-la como \emph{proxy} informativa do componente latente de incerteza geopolítica.

Ambas as séries são $I(1)$ e cointegradas com $r=1$. Ao estimar um VEC com uma defasagem ($p=1$) e constante dentro do espaço de cointegração, o modelo reporta a seguinte combinação de longo prazo entre os níveis de GPR e micromobilidade:
\begin{equation}
\underbrace{\beta' y_{t-1}}_{\mathrm{ECT}_{t-1}}
= \mathrm{GPR}_{t-1} - 0{,}091\,\mathrm{Veh}_{t-1} - 2{,}319,
\end{equation}
onde $\mathrm{ECT}_{t-1}$ (``\emph{error-correction term}'') é o desvio em relação ao equilíbrio de longo prazo: no equilíbrio, $\mathrm{ECT}_{t-1}=0$\footnote{A escrita acima segue a convenção de normalizar o coeficiente da GPR em $1$; essa escolha de escala não altera a interpretação econômica.}. Essa relação é positiva porque, isolando o equilíbrio ($\mathrm{ECT}_{t-1}=0$), obtemos
\begin{equation}
\mathrm{GPR}^\star_{t-1} \;=\; 0{,}091\,\mathrm{Veh}_{t-1} \;+\; 2{,}319, \label{eq:gpr_equilibrio}
\end{equation}
isto é, níveis mais altos de micromobilidade local estão associados a um nível de GPR de equilíbrio mais alto; e estável porque a combinação $\beta'y_{t-1}$ é estacionária por construção (cointegração), de modo que desvios em relação a essa relação tendem a se dissipar ao longo do tempo.

O VEC é uma reparametrização do VAR em que modelamos diferenças e incluímos o termo de correção de erro. No nosso bivariado $y_t=[\mathrm{GPR}_t,\ \mathrm{Veh}_t]'$, a forma geral
\begin{equation}
\Delta y_t \;=\; \alpha\,\beta' y_{t-1} \;+\; \Gamma\,\Delta y_{t-1} \;+\; \varepsilon_t
\end{equation}
se traduz em duas equações (uma para cada variável), nas quais o vetor $\beta$ aparece dentro do $\mathrm{ECT}_{t-1}$ (a relação de longo prazo) e o vetor $\alpha$ multiplica esse erro (as velocidades de ajuste de curto prazo):
\begin{equation}
\begin{aligned}
\Delta \mathrm{GPR}_t \;&=\; \underbrace{\alpha_{\mathrm{GPR}}}_{-0{,}378^{***}}\,
\underbrace{\big(\mathrm{GPR}_{t-1} - 0{,}091\,\mathrm{Veh}_{t-1} - 2{,}319\big)}_{\mathrm{ECT}_{t-1}}
\;+\; \gamma_{11}\Delta \mathrm{GPR}_{t-1} \;+\; \gamma_{12}\Delta \mathrm{Veh}_{t-1}
\;+\; \varepsilon_{1t},\\[2mm]
\Delta \mathrm{Veh}_t \;&=\; \underbrace{\alpha_{\mathrm{Veh}}}_{0{,}161\ \text{(n.s.)}}\,
\underbrace{\big(\mathrm{GPR}_{t-1} - 0{,}091\,\mathrm{Veh}_{t-1} - 2{,}319\big)}_{\mathrm{ECT}_{t-1}}
\;+\; \gamma_{21}\Delta \mathrm{GPR}_{t-1} \;+\; \gamma_{22}\Delta \mathrm{Veh}_{t-1}
\;+\; \varepsilon_{2t}.
\end{aligned}
\end{equation}

\medskip
A intuição operacional é simples. Suponha, na Eq.~(\ref{eq:gpr_equilibrio}), $\mathrm{Veh}_{t-1}=10$. O equilíbrio de longo prazo pediria então $\mathrm{GPR}^{\star}_{t-1}=0{,}091\times 10 + 2{,}319=3{,}229$. Se observamos $\mathrm{GPR}_{t-1}=4$, então o desvio em relação ao equilíbrio de longo prazo é positivo ($\mathrm{ECT}_{t-1}=4-3{,}229=+0{,}771$) e, como $\alpha_{\mathrm{GPR}}<0$, a equação implica $\Delta \mathrm{GPR}_t<0$: a GPR cai no período seguinte para voltar ao equilíbrio. Já a micromobilidade tem $\alpha_{\mathrm{Veh}}$ não significativo, sugerindo que ela não é a variável que ``puxa'' o sistema de volta. Ela captura o estado latente, mas não determina a correção.

\begin{figure}[H]
\centering
\begin{subfigure}[t]{0.48\textwidth}
  \centering
  \includegraphics[width=\linewidth]{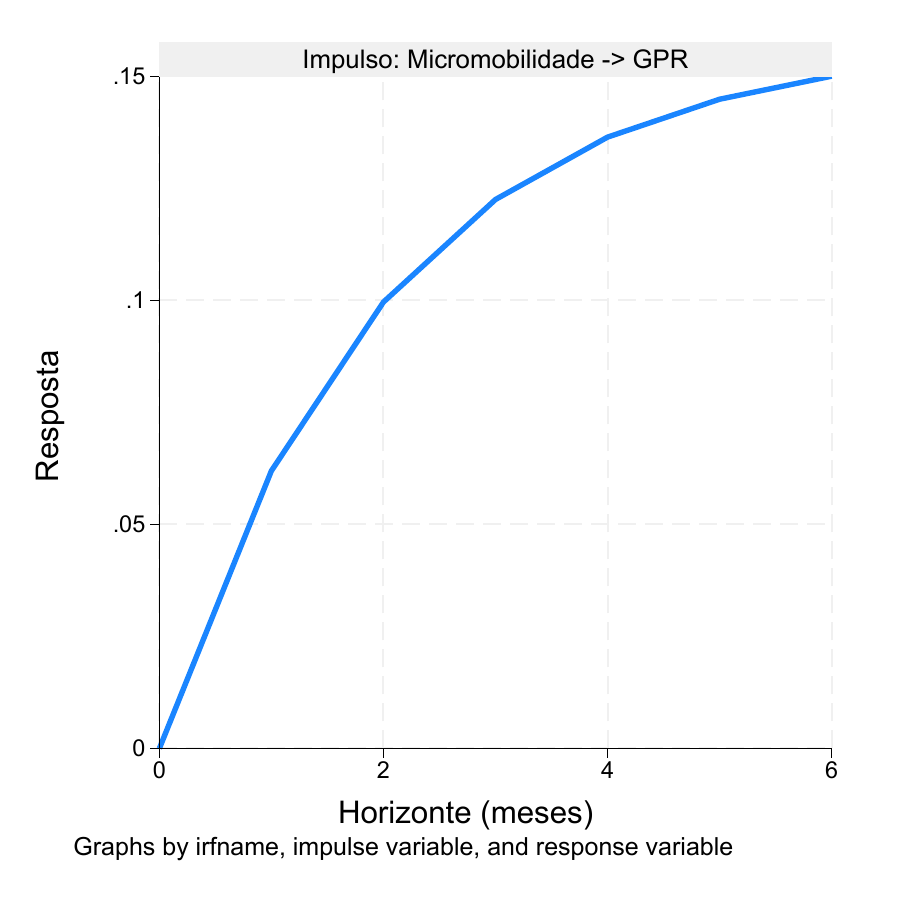}
  \caption{Impulso: Micromobilidade $\to$ GPR}
  \label{fig:irf_1}
\end{subfigure}\hfill
\begin{subfigure}[t]{0.48\textwidth}
  \centering
  \includegraphics[width=\linewidth]{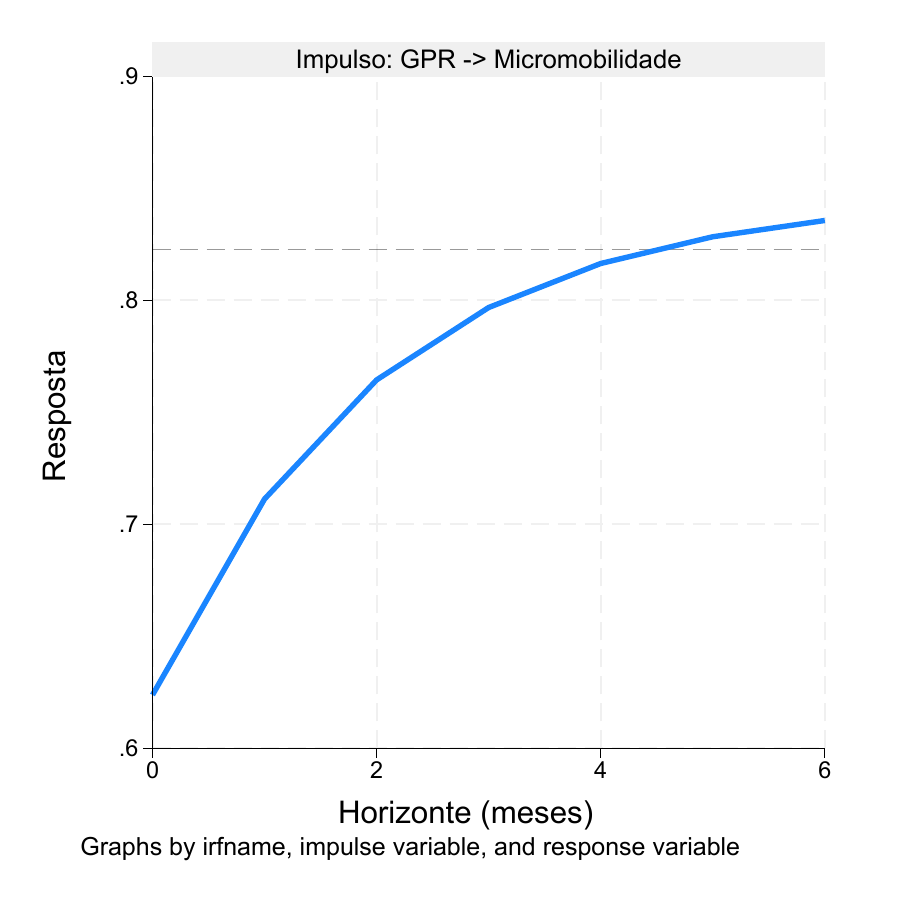}
  \caption{Impulso: GPR $\to$ Micromobilidade}
  \label{fig:irf_2}
\end{subfigure}

\vspace{0.6em}

\begin{subfigure}[t]{0.48\textwidth}
  \centering
  \includegraphics[width=\linewidth]{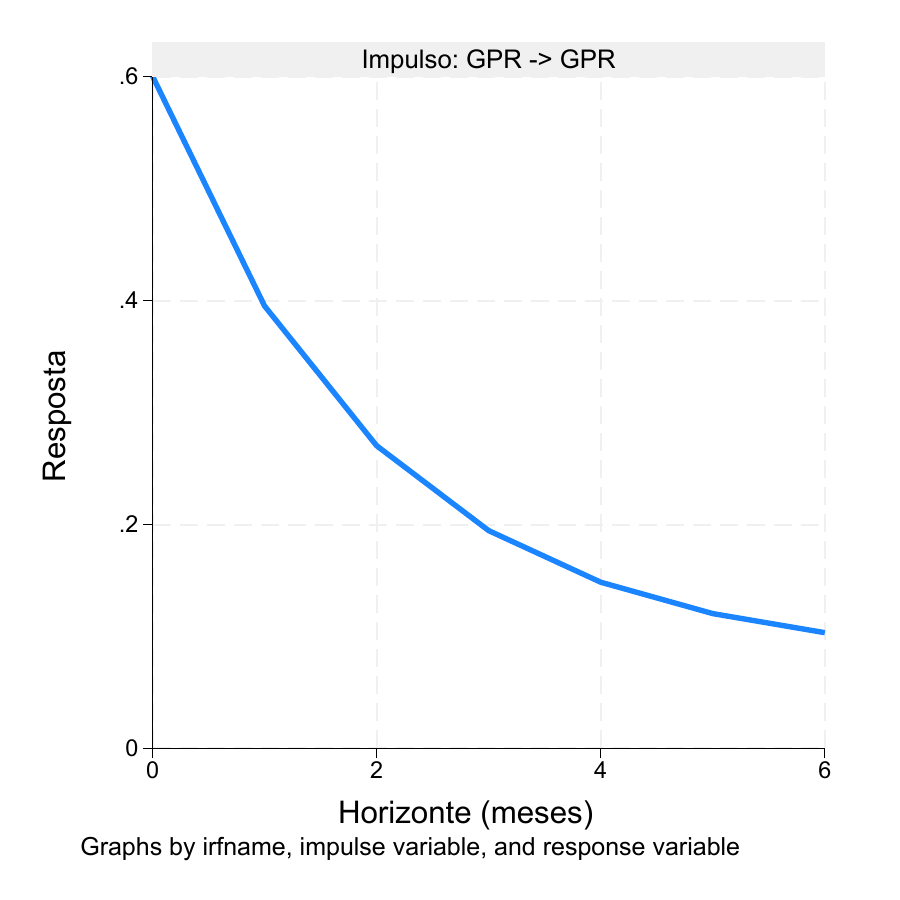}
  \caption{Impulso: GPR $\to$ GPR}
  \label{fig:irf_3}
\end{subfigure}\hfill
\begin{subfigure}[t]{0.48\textwidth}
  \centering
  \includegraphics[width=\linewidth]{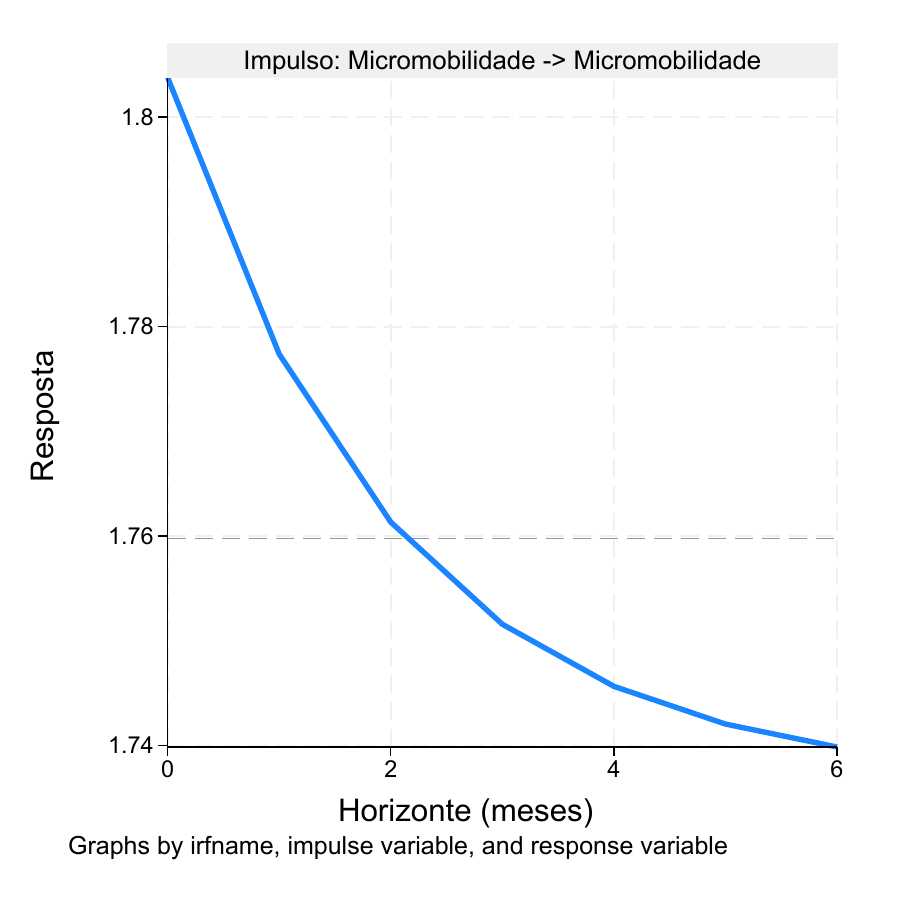}
  \caption{Impulso: Micromobilidade $\to$ Micromobilidade}
  \label{fig:irf_4}
\end{subfigure}
\caption{Funções Impulso--Resposta (VEC bivariado)}
\label{fig:irf_vec_panel}
\end{figure}

Essa leitura casa com a interpretação de \emph{proxy}. Quando o fator latente de tensão geopolítica $C_t$ aumenta, ele eleva tanto o GPR (medida agregada nacional) quanto a micromobilidade local; por isso, as duas séries co-movem ao responder ao mesmo impulso subjacente. No VEC, isso se manifesta como um vínculo de longo prazo positivo entre os níveis e como correção do desvio principalmente pela GPR (velocidade de ajuste significativa e negativa). Por consequência, os choques cruzados tendem a ser positivos, pois um aumento em uma série vem acompanhado de aumento na outra porque ambas são empurradas pelo mesmo motor latente. Já as respostas próprias tendem a reverter porque, após o choque inicial, o mecanismo de correção de erro força a variável que se afastou do equilíbrio a mover-se na direção oposta nos períodos seguintes, dissipando gradualmente o impacto. Este é exatamente o padrão observado na Fig.~\ref{fig:irf_vec_panel}.

Depois de toda essa discussão, será que a nossa \emph{proxy} passa pelo crivo das diretrizes definidas anteriormente? Interpretamos a micromobilidade em Arlington (\(\mathrm{Veh}_t\)) como \emph{proxy} imperfeita para o fator latente de tensão geopolítica (\(C_t\)) que move o risco geopolítico agregado (\(\mathrm{GPR}_t\)).

\paragraph{Relevância.}
A combinação estacionária estimada pelo VEC (\(r=1\)) foi
\[
\mathrm{ECT}_{t-1}=\mathrm{GPR}_{t-1}-0{,}091\,\mathrm{Veh}_{t-1}-2{,}319,
\]
ou seja, quanto maior a micromobilidade local, maior o nível de GPR consistente com o equilíbrio.

\paragraph{Suficiência condicional.}
Na linguagem da projeção \(C=\delta_0+\delta_X X+\delta_P P+V\), a hipótese chave é que, dado o valor da \emph{proxy} \(P\) (a micromobilidade), outras variáveis \(X\) não acrescentem informação adicional relevante sobre \(C\), isto é, \(\delta_X=0\). No nosso exercício bivariado, essa suposição equivale a dizer que a flutuação local de micromobilidade ao redor do Pentágono já carrega, por si, o sinal do ``ímpeto geopolítico'' que também afeta o GPR; é uma hipótese de identificação, não um resultado estatístico.

\paragraph{Exogeneidade.}
Exige-se que a \emph{proxy} afete \(\mathrm{GPR}_t\) apenas por meio de sua correlação com \(C_t\) (formulado como \(\mathbb{E}[U\mid X,P]=0\) no modelo estrutural). Em termos econômicos, a micromobilidade ``registra'' o aquecimento operacional local quando a agenda de defesa se intensifica, mas não deveria haver um canal direto e autônomo para alterar o risco geopolítico nacional fora desse mecanismo comum. Isso equivaleria a dizer que um aumento de patinetes na calçada, por si só, faria o Conselho de Segurança Nacional elevar o nível de alerta, como se decisões estratégicas fossem tomadas olhando o fluxo de scooters. Essa leitura não faz sentido: o fluxo de micromobilidade só cresce porque a atividade ligada à defesa se intensifica (o fator latente), não porque ele determine o risco geopolítico nacional.

\paragraph{Estabilidade.}
O fato de as séries serem cointegradas sinaliza uma relação de longo prazo estável no período analisado (2019--2024): a mesma combinação linear retorna ao equilíbrio após choques, o que é a noção de ``ponte'' estável entre a medida local (micromobilidade) e o risco agregado (GPR).

\paragraph{Conclusão da proposta.}
A ideia de usar micromobilidade em Arlington como \emph{proxy} operacional da tensão geopolítica é economicamente motivada e encontra respaldo nas evidências: cointegração com o GPR, coeficiente de longo prazo positivo e respostas cruzadas coerentes. No enquadramento de \emph{proxy}, o exercício atende à relevância (a micromobilidade entra na relação de equilíbrio), é compatível com suficiência condicional (postula que o sinal local captura a informação central do fator latente), respeita exogeneidade como narrativa causal (a \emph{proxy} apenas registra o mesmo impulso que move o GPR) e sugere estabilidade via cointegração no período analisado. Em suma, a micromobilidade atua como um ``termômetro'' local do mesmo processo latente que alimenta o risco geopolítico nacional: não causa o risco, mas o acompanha de perto e, por isso, permite inferir o invisível a partir do observável.

Isso nos leva à conclusão fundamental. A validade de uma variável \emph{proxy} não é, em última instância, um veredito estatístico, mas sim o resultado de um argumento bem construído. As quatro condições que exploramos (relevância, suficiência condicional, exogeneidade e estabilidade) não são testadas com um clique, mas defendidas com base na teoria econômica e no conhecimento do mundo real. O caso das pizzarias do Pentágono, embora anedótico, ilustra bem essa arte: combinar criatividade para encontrar medidas engenhosas com rigor analítico para compreender e comunicar suas limitações. Medir o que não é diretamente mensurável é, portanto, construir uma ponte de hipóteses defensáveis entre o mundo que vemos nos dados e a complexa realidade que desejamos entender.

\bibliographystyle{apalike}
\bibliography{sample}

\end{document}